\documentclass{article}
\usepackage[utf8]{inputenc}
\usepackage{authblk}
\usepackage{setspace}
\usepackage[margin=1.25in]{geometry}
\usepackage{graphicx}
\graphicspath{{./figures/}}
\usepackage{amsmath}
\usepackage{amssymb}
\usepackage{lineno}
\usepackage{microtype}
\usepackage{xurl}
\usepackage[hidelinks]{hyperref}

\usepackage[
  backend=bibtex,
  style=nejm,
  citestyle=numeric-comp,
  sorting=none
]{biblatex}
\title{Constraint-Traceable Liquid-Nitrogen Dewar Design for Low-Noise HTS-SQUID Magnetometry}

\author[1,$\dagger$]{Xinmin Shi}
\author[1,$\dagger$, *]{Bingke Xiang}
\author[1,$\dagger$]{Lingtong Hou}
\author[2]{Wanjuan Tang}
\author[2]{Geming Zhang}
\author[2]{Shiqun Liu}
\author[1]{Ruonan Wang}
\author[1]{Yibo Wang}
\author[1]{Zhiqiang Cao}
\author[3,*]{Xueshen~Wang}
\author[1,2,4]{Xueying~Zhang}
\author[1,2,4,*]{Xiaoyang Lin}

\affil[1]{State Key Laboratory of Spintronics, Hangzhou International Innovation Institute, Beihang University, Hangzhou 311115, China}
\affil[2]{Fert Beijing Institute, School of Integrated Circuit Science and Engineering, Beihang University, Beijing 100191, China}
\affil[3]{National Institute of Metrology, Beijing 100029, China}
\affil[4]{Truth Instruments Co. Ltd., Qingdao 266100, China}
\affil[*]{Address correspondence to:
\mbox{\href{mailto:xiangbk@buaa.edu.cn}{\textnormal{xiangbk@buaa.edu.cn}}},
\mbox{\href{mailto:wangxs@nim.ac.cn}{\textnormal{wangxs@nim.ac.cn}}}, and
\mbox{\href{mailto:XYLin@buaa.edu.cn}{\textnormal{XYLin@buaa.edu.cn}}}}
\affil[$\dagger$]{These authors contributed equally to this work.}

\date{}
\begin{document}

\maketitle

\begin{abstract}
Compact high-temperature superconducting quantum interference device (HTS-SQUID) magnetometers require a liquid-nitrogen Dewar that balances operating duration against instrument-envelope and mass constraints. Here we introduced a constraint-traceable design workflow in which body radius and neck-length fraction were scanned under fixed outer-envelope volume and geometry-based mass limits. The longest-duration feasible grid point lay adjacent to the 3.85 kg and 550 mm constraint intersection, predicting 224 h hold time from a body-only initial fill to a residual liquid depth of 10 mm. A reference Dewar was experimentally monitored for 137.3 h, consuming 1.22 L liquid-nitrogen of its initial inventory. A thermal model calibrated over the first 40 h reproduced the remaining 97.3 h. Device measurements at liquid-nitrogen temperature showed a superconducting transition and Fraunhofer-like junction response, a maximum voltage-modulation depth of 36.2 $\mu$V, and a median noise level of 41.5 fT Hz$^{-1/2}$ from 100 to 1000 Hz. These results establish a reproducible method for selecting Dewar geometry under coupled constraints and demonstrate compatibility with long-endurance and low-noise HTS-SQUID magnetometer systems.
\end{abstract}


\section{Introduction}

Superconducting quantum interference devices (SQUIDs) are among the most
sensitive magnetic-flux detectors and have enabled precision measurements
across a broad range of scientific and technological fields
\cite{Koelle1999HTSSQUIDReview,clarke2004squid,fagaly2006squid,wang2026htsreview,kirtley2010scanning,Xiang2024EMreview}.
High-temperature superconducting SQUIDs (HTS-SQUIDs), particularly devices
based on cuprate superconductors, can operate near the boiling point of liquid
nitrogen. Compared with conventional low-temperature SQUID
systems that generally require liquid-helium cooling, operation near 77 K
reduces the complexity and operating cost of the cryogenic subsystem
\cite{Koelle1999HTSSQUIDReview,clarke2004squid}.

The availability of liquid-nitrogen cooling has supported the development of
HTS-SQUID instruments for biomagnetic measurements
\cite{korber2016biomagnetism,faley2017Bio}, nondestructive evaluation
\cite{Lucia1997PortableCryostat,braginski2000nde,tanaka2020,sun2023,wang2024} and geophysical and mineral exploration
\cite{foley1999field,hato2013development,wu2021}. These applications frequently require the sensor
to operate outside a conventional laboratory environment, where instrument
volume, mass, transportability, sensor-to-sample spacing, and unattended
measurement duration become system-level design requirements. A liquid-nitrogen Dewar is not merely a passive cryogen reservoir: its geometry
simultaneously determines the stored liquid inventory, conductive heat-flow
paths, radiative surface area, mechanical envelope, sensor accessibility, and
operating duration. Increasing liquid volume may extend the nominal hold time,
but it can also increase instrument dimensions and mass or compromise the
working distance between the SQUID and the measured object. Dewar development
must therefore address coupled geometric, thermal, and instrument-level
constraints rather than maximize cryogen volume alone.

Several studies have developed specialized cryostats for portable HTS-SQUID
systems\cite{hato2013development,faley2017,pfeiffer2020}. Lucia et al.\ demonstrated an HTS-SQUID system incorporating a
portable cryostat for eddy-current nondestructive evaluation
\cite{Lucia1997PortableCryostat}. Hato and Tanabe analyzed conductive and radiative
heat loads in compact glass Dewars and reported cooling for more than 100 h
using approximately 0.8 L of liquid nitrogen
\cite{Hato2018LongCooling}. A temperature-controlled portable cryostat was
subsequently developed to adjust the operating temperature and slew-rate
performance of an HTS-SQUID
\cite{hato2019portable}. Long-hold-time cryogenic vessels have also been
reported for other precision instruments, demonstrating the importance of
co-designing the cryogen inventory, structural supports, thermal shielding,
and overall vessel geometry
\cite{fixsen2001lightweight}. More generally, reductions in radiation and
solid conduction are commonly pursued through low-conductivity structural
paths, vacuum insulation, radiation shields, and multilayer insulation
\cite{kropschot1961mli}.

These studies provide important cryostat implementations and thermal-design
principles, but they primarily evaluate particular vessel structures or a
limited set of candidate geometries. Such comparisons do not, by themselves,
provide a constraint-explicit rule for selecting dimensions when instrument
volume, total length, structural mass, and operating duration are imposed
simultaneously. In a compact HTS-SQUID instrument, the preferred geometry may
therefore be determined not by an unconstrained thermal optimum, but by the
intersection of several active engineering constraints. A design workflow
should consequently identify not only the predicted hold time, but also the
feasible region, the active constraint boundaries, and the design margin
available around the selected point.

Here, we formulate the design of a compact liquid-nitrogen Dewar as a
constraint-traceable geometric selection problem. The inner radius of the body and
the neck-length fraction are scanned under declared limits of outer-envelope volume and geometry-based mass. We address three interrelated questions: (i) how to establish a constraint-traceable design map that delineates the feasible region, identifies the active constraint boundaries, and quantifies the trade-off between hold time and design margin around the selected geometry; (ii) how to validate the thermal model using a chronological calibration–holdout–extrapolation protocol; and (iii) whether the resulting liquid-nitrogen platform can support low-noise HTS-SQUID operation, as verified at the device level through Josephson-junction characterization, SQUID voltage–flux modulation, and noise measurements.
The resulting workflow provides an auditable connection between engineering requirements, geometric selection, thermal prediction, and instrument-level operation.

\section{Materials and Methods}

\subsection{Experimental Design}

The study comprised four linked evidence modules. Figure~\ref{fig:1} defines the Dewar geometry, heat-flow model, constraints, and deterministic design scan. Figure~\ref{fig:2} compares the model with a continuous weighing record from a reference G10 Dewar and displays the model-assigned conductive and radiative heat loads. Figure~\ref{fig:3} characterizes the superconducting transition and field-dependent critical current of a Josephson junction. Figure~\ref{fig:4} reports SQUID voltage-flux modulation and a separately acquired noise spectrum at 77 K. The electrical measurements assess compatibility with HTS-SQUID operation in the liquid-nitrogen Dewar.

\subsection{Constrained Geometry}

The independent design variables were the body liquid inner radius $R$ and the neck-length fraction $r=L_{\mathrm{neck}}/L$ (Fig.~\ref{fig:1}a). For a fixed outer-envelope volume $V_{\mathrm{outer}}=7.1\times10^{-3}$ m$^3$,

\begin{equation}
L(R)=\frac{V_{\mathrm{outer}}}{\pi(R+\delta_R)^2},
\qquad \delta_R=25~\mathrm{mm},
\label{eq:length}
\end{equation}

\begin{equation}
L_{\mathrm{neck}}=rL,\qquad
L_{\mathrm{body}}=L-L_{\mathrm{neck}}-\delta_L,
\qquad \delta_L=8~\mathrm{mm}.
\label{eq:geometry}
\end{equation}

The offset $\delta_L$ reproduces the reference geometry $R=39$ mm, $L_{\mathrm{neck}}=200$ mm, and $L_{\mathrm{body}}=344$ mm. The body and neck G10 walls were each 2 mm thick; the neck inner diameter was 55 mm. Initial liquid was assigned only to the body and the neck was not initially filled. The body fill fraction, $f_{\mathrm{fill}}=1$, was transferred to each design point:

\begin{equation}
h_0=f_{\mathrm{fill}}L_{\mathrm{body}},\qquad
m_0=\rho_{\mathrm{LN2}}\pi R^2h_0,
\label{eq:initialmass}
\end{equation}

with $\rho_{\mathrm{LN2}}=807$ kg m$^{-3}$. The endpoint was a residual body liquid depth of 10 mm rather than complete depletion.

The geometry-based mass proxy was

\begin{equation}
M_{\mathrm{model}}=m_0+\rho_{\mathrm{G10}}\sum_i V_{\mathrm{G10},i},
\label{eq:massproxy}
\end{equation}

where $\rho_{\mathrm{G10}}=1990$ kg m$^{-3}$ and the summed volumes represented the parameterized body wall, neck wall, bottom, shoulder, outer wall, and outer end plates. It did not include multilayer insulation, foam, adhesive, fittings, valves, wiring, or other assembly hardware and therefore was not an as-built mass. The feasible region was defined by $M_{\mathrm{model}}\leq3.85$ kg and $L\leq550$ mm. The two constraint limits originate from instrument-level requirements of the target compact HTS-SQUID magnetometer platform. The overall-height limit was imposed by the restricted vertical clearance available in representative measurement environments. In biomagnetic applications such as magnetocardiography, for example, the Dewar must be accommodated within the geometry defined by the patient bed, the subject position, and the supporting structure. An excessively tall Dewar would therefore complicate system integration and limit compatibility with practical measurement sites. The mass limit was set by the requirements for one-person handling during field deployment and by the payload capacity of the instrument mount.


The scanned ranges were chosen to bracket the practical design window. Hold time was calculated throughout the grid, while the constraint boundaries were interpolated independently. The exact boundary intersection served as a continuous numerical check of the adjacent resolution-limited grid maximum.

\subsection{Thermal Model}

The model contained a series G10 conduction path and liquid-level-dependent sidewall and bottom radiation (Fig.~\ref{fig:1}a). The temperature-integrated G10 conductivity was calculated from the NIST G-10 CR warp-direction fit:

\begin{equation}
K_{\mathrm{int}}=\int_{77}^{295.15}k_{\mathrm{G10}}(T)\,\mathrm{d}T
=139.08~\mathrm{W\,m^{-1}}.
\label{eq:kint}
\end{equation}

For neck and body-wall axial areas $A_{\mathrm{neck}}$ and $A_{\mathrm{body}}$, respectively,

\begin{equation}
Q_{\mathrm{cond}}(h)=
\frac{K_{\mathrm{int}}}
{L_{\mathrm{neck}}/A_{\mathrm{neck}}
+(L_{\mathrm{body}}-h)/A_{\mathrm{body}}}.
\label{eq:qcond}
\end{equation}

Radiative terms were

\begin{equation}
Q_{\mathrm{rad,side}}(h)=
\epsilon_{\mathrm{side}}\sigma(2\pi Rh)(T_h^4-T_c^4),
\label{eq:qradside}
\end{equation}

\begin{equation}
Q_{\mathrm{rad,bottom}}=
\epsilon_{\mathrm{bottom}}\sigma(\pi R^2)(T_h^4-T_c^4),
\label{eq:qradbottom}
\end{equation}

where $T_{\mathrm{h}}=295.15~\mathrm{K}$, $T_{\mathrm{c}}=77~\mathrm{K}$, and $\sigma=5.67\times10^{-8}~\mathrm{W\,m^{-2}\,K^{-4}}$. The nominal effective emissivities of the sidewall and bottom were set as $\epsilon_{\mathrm{bottom}}/\epsilon_{\mathrm{side}}=2.5$. These unequal values reflect the deliberately asymmetric multilayer-insulation (MLI) configuration of the Dewar. The effective thermal performance of MLI depends strongly on the number and packing density of the reflective layers. The sidewall provides sufficient radial clearance to accommodate more MLI layers, whereas fewer layers are used at the bottom to minimize the SQUID-to-sample standoff. The bottom therefore exhibits a systematically higher effective radiative coupling to the warm environment than the sidewall. Because this relative insulation configuration is fixed by the hardware geometry, the ratio $\epsilon_{\mathrm{bottom}}/\epsilon_{\mathrm{side}}$ was held constant at 2.5, while a single common scale factor was introduced to account for system-level deviations from the nominal MLI performance. The common scale factor was calibrated using only the first 40~h of the weighing record, yielding $\epsilon_{\mathrm{side}}\approx0.008$ and $\epsilon_{\mathrm{bottom}}\approx0.021$.

The total modeled heat load and mass-loss equation were

\begin{equation}
Q_{\mathrm{total}}(h)=Q_{\mathrm{cond}}(h)
+Q_{\mathrm{rad,side}}(h)+Q_{\mathrm{rad,bottom}},
\label{eq:qtotal}
\end{equation}

\begin{equation}
\frac{\mathrm{d}m}{\mathrm{d}t}
=-\frac{Q_{\mathrm{total}}(m)}{L_v},
\qquad L_v=199000~\mathrm{J\,kg^{-1}}.
\label{eq:massloss}
\end{equation}

The predicted duration to the prescribed endpoint was

\begin{equation}
t_{10}=
\frac{\rho_{\mathrm{LN2}}\pi R^2L_v}{3600}
\int_{0.010}^{h_0}\frac{\mathrm{d}h}
{Q_{\mathrm{total}}(h;R,r)}.
\label{eq:holdtime}
\end{equation}

Nitrogen density, latent heat, and the standard-pressure boiling-point anchor were treated as model inputs based on tabulated properties rather than measurements of the local bath. The model did not explicitly resolve vapor cooling, the foam plug, residual gas conduction, neck-opening radiation, joints, adhesive layers, or wiring; their combined influence can be confounded with the fitted effective parameters.

\subsection{Hold-Time Measurement}

The Dewar was placed on an electronic weighing scale for continuous mass-loss monitoring at an ambient temperature of approximately $22\,^{\circ}\mathrm{C}$ ($295.15~\mathrm{K}$). A custom-made empty-Dewar mass of 2.558 kg was subtracted to obtain the liquid-nitrogen inventory. The inventory decreased from 1.318 kg at the start to 0.330 kg when recording ended. The measurement did not reach the 10 mm endpoint, whose calculated residual mass for the reference $R=39$ mm body was 0.038 kg.

The effective emissivity scale was determined by one-dimensional minimization against data acquired during the first 40 h. All subsequent observations, extending from 40 h to the end of the 137.3 h measurement, were withheld from parameter estimation and used exclusively for chronological validation.

Agreement was summarized as

\begin{equation}
\mathrm{RMSE}=
\sqrt{\frac{1}{n}\sum_{i=1}^{n}
\left[m_{\mathrm{model}}(t_i)-m_{\mathrm{meas}}(t_i)\right]^2}.
\label{eq:rmse}
\end{equation}

No inferential confidence interval was assigned because the record represents one endurance run. The model was then integrated beyond the measurement window to the prescribed 10 mm endpoint.

\subsection{Josephson-junction characterization}
\label{sec:jj_characterization}

The resistive transition of the Josephson junctions was measured by phase-sensitive lock-in detection over the temperature range $77$--$100~\mathrm{K}$. The thermally broadened low-resistance tail of the transition was analyzed using the Ambegaokar--Halperin description of a Josephson junction dominated by fluctuation \cite{AmbegaokarHalperin1969}. The fit was restricted to $81$--$87~\mathrm{K}$, and the transition temperature was constrained to $T_{\mathrm{c},90}$, defined by the 90\% resistance criterion. The fitted temperature exponent was treated as a phenomenological measure of transition broadening.

Field-dependent current--voltage characteristics were acquired during two nominal-field sweeps performed in opposite directions. 
The positive- and negative-bias branches were fitted independently using the overdamped limit of the resistively shunted junction model \cite{Stewart1968a,McCumber1968} to extract the critical current $I_c$. Because of hysteresis between the increasing and decreasing field sweeps, each branch was shifted to align the maximum $I_c$ with zero field. The negative-field half of the increasing sweep and the positive-field half of the decreasing sweep were then combined to form the complete field-response map.

\subsection{SQUID Modulation and Noise}

The HTS SQUID employed in the present Dewar evaluation was fabricated using our recently developed morphology-guided process \cite{xiang2026}. Voltage-flux traces were acquired at bias currents of 40, 45, 50, 55, 60, 65, and 70 $\mu$A. The modulation period was a relative period-normalized coordinate. Modulation depth was the median peak-to-peak voltage over four complete displayed cycles.

The voltage-noise spectrum was acquired using a dynamic signal analyzer and converted to flux noise using the calibrated SQUID flux-to-voltage response. The broadband noise level was quantified as the median spectral density over 100--1000 Hz. The magnetic-field noise was obtained from the flux noise using an effective area of $A_{\rm eff}=0.35~\mathrm{mm^2}$. The modulation and noise measurements were performed separately.

\section{Results}

\subsection{Constraints Set the Design Point}

The design scan revealed a clear transition between length-limited and mass-limited geometries (Fig. 1b). The fixed outer volume made total length decrease monotonically with body radius, so the 550 mm boundary occurred at $R=39$ mm. At that radius, the 3.85 kg mass-proxy boundary intersected at $r=0.361$. The Hold-time contours increased toward this intersection from within the feasible region. The continuous intersection predicted $t_{10}=224$ h. 

\begin{figure}[!p]
    \centering
    \includegraphics[width=\textwidth]{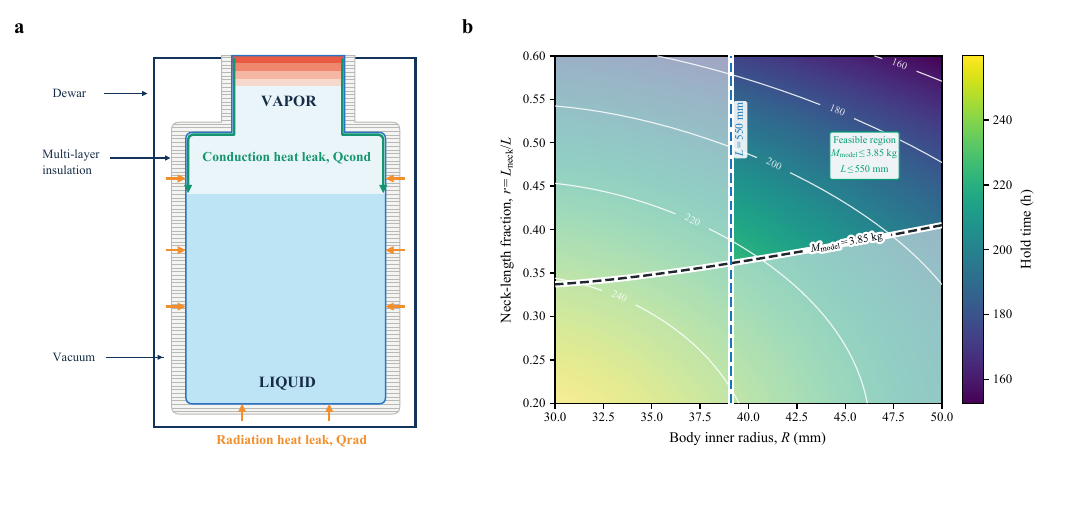}
    \caption{\textbf{Thermal model and constrained hold-time design space of the liquid-nitrogen Dewar.} \textbf{a,} Schematic of the G10 Dewar and the modeled conductive ($Q_{\mathrm{cond}}$) and radiative ($Q_{\mathrm{rad}}$) heat-load paths. \textbf{b,} Calculated liquid-nitrogen hold time, from a body-only initial fill to a residual liquid depth of 10 mm in the body, as a function of body inner radius $R$ and neck-length fraction $r=L_{\mathrm{neck}}/L$. White contours indicate hold time; the black and blue dashed lines denote the $M_{\mathrm{model}}=3.85$ kg and $L=550$ mm constraints, respectively. The unfrosted region satisfies both constraints. $M_{\mathrm{model}}$ is a geometry-based mass proxy.}
    \label{fig:1}
\end{figure}

Thus, the longest-duration feasible design was not an unconstrained thermal optimum but was set directly by the simultaneous use of the available height and mass budgets. Both permitted length and geometry-based mass are nearly fully used, and the exact intersection contains no fabrication or assembly reserve. The unfrosted region of Fig.~\ref{fig:1}b exposes the nearby alternatives that trade a small reduction in predicted duration for greater margin. Because the unconstrained maximum lies at the boundary of the scanned domain, the result is limited to the declared design space.

\subsection{Model Tracks Measured Inventory}

\begin{figure}[!p]
    \centering
    \includegraphics[width=\textwidth]{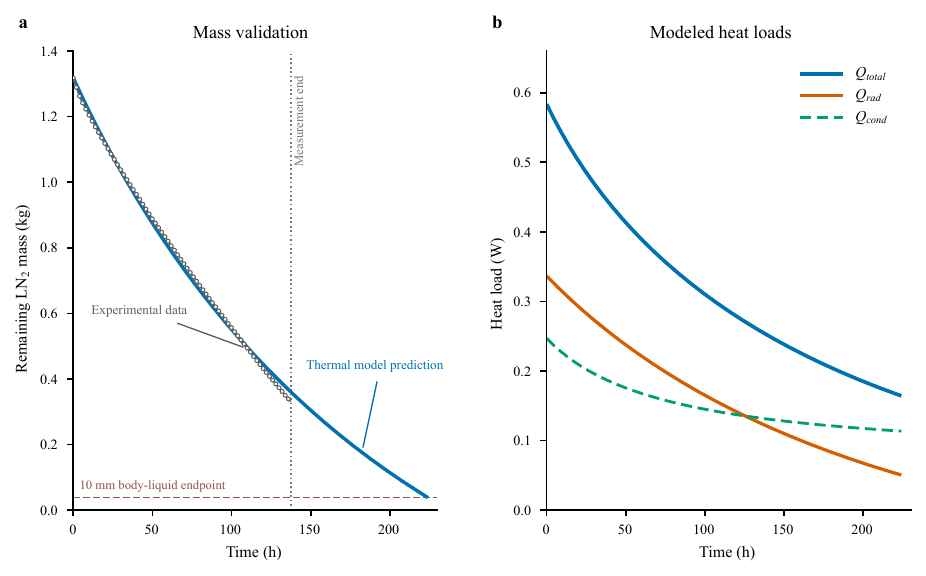}
    \caption{\textbf{Hold-time validation and evolution of heat loads in the liquid-nitrogen Dewar.} \textbf{a,} Measured remaining liquid-nitrogen mass (open grey circles) and thermal-model prediction (blue line). The vertical dotted line marks the end of the measurement, while the horizontal dashed line indicates the defined endpoint of a 10 mm liquid level in the Dewar body, predicted at approximately 224 h. \textbf{b,} Model-calculated total heat load ($Q_{\mathrm{total}}$) and its radiative ($Q_{\mathrm{rad}}$) and conductive ($Q_{\mathrm{cond}}$) components, all of which decrease as the liquid level falls. The approximately 224 h endpoint is a model extrapolation beyond the measured interval.}
    \label{fig:2}
\end{figure}

The reference-Dewar inventory declined from 1.318 to 0.330 kg over 137.3 h (Fig.~\ref{fig:2}a). After calibration over 0--40 h, the model reproduced the 40--137.3 h holdout with an RMSE of 0.014 kg. The measured average effective heat load over the observed interval was 0.397 W, compared with 0.385 W from the model.

Extending the reference-geometry model beyond the observed record reached the prescribed 10 mm body liquid depth at 224 h. Thus 137 h is the measured retention interval, whereas approximately 224 h is a model-defined endpoint prediction. The model-assigned total heat load decreased from 0.584 W initially to 0.254 W at 137 h and 0.164 W at the extrapolated endpoint (Fig.~\ref{fig:2}b). In the model, the decline follows both the increasing dry-wall conduction length and the shrinking liquid-cooled sidewall area. The conductive and radiative components were not measured independently.

Because the validation record was obtained from a single reference Dewar, the agreement quantified here evaluates the model within one vessel and does not establish vessel-to-vessel reproducibility. Accordingly, the fitted effective-emissivity parameters and the 224 h endpoint prediction should be regarded as specific to the present reference geometry and model assumptions.

\subsection{Junction Shows Josephson Response}

\begin{figure}[!p]
    \centering
    \includegraphics[width=\textwidth]{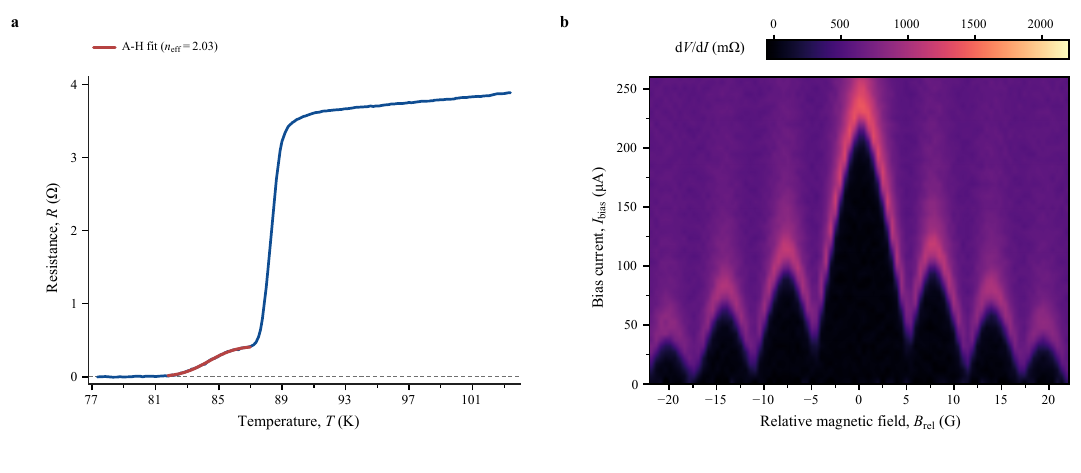}
    \caption{\textbf{Superconducting transition and magnetic-field response of a Josephson junction.} \textbf{a,} Resistance-temperature curve with an Ambegaokar-Halperin fit to the low-resistance foot. \textbf{b,} Differential-resistance map as functions of magnetic field and bias current. The displayed field is an offset-corrected nominal coordinate.}
    \label{fig:3}
\end{figure}

The processed resistance-temperature curve exhibited a finite-width superconducting transition, with $T_{10}=85.746$ K, $T_{50}=88.240$ K, and $T_{90}=89.001$ K on the standard-pressure-anchored temperature axis, giving $\Delta T_{c,10-90}=3.255$ K (Fig.~\ref{fig:3}a). The selected 81–87 K interval was described by the conditional Ambegaokar–Halperin model, yielding $n_{\mathrm{eff}}=2.03$. Within the adopted parameterization, $n_{\mathrm{eff}}$ determines how rapidly the effective Josephson coupling decreases on approaching the transition. The observed low-resistance tail is therefore consistent with thermally broadened Josephson transport.

The field-dependent transport provides an independent signature of Josephson coupling (Fig. 3b). The offset-corrected differential-resistance map displayed a central maximum critical current of 222.50 $\mu$A and two resolved side lobes (Fig.~\ref{fig:3}b). The strong magnetic modulation of critical current and the persistence of side lobes are characteristic of coherent Josephson interference across the grain boundary. The response nevertheless deviates from an ideal sinc pattern, indicating that the measured junction cannot be represented by an idealized uniform-current. A field-history dependence was observed between the two sweep directions. After correcting these offsets and combining the corresponding half-sweeps, the interference structure remained well resolved. The field-history offset is consistent with screening and flux-trapping effects known in planar HTS Josephson junctions, although the present measurement does not uniquely identify the underlying mechanism. 

Together, the thermally broadened resistive transition and field-modulated critical current provide complementary evidence that the junction retains Josephson weak-link operation under liquid-nitrogen conditions.

\subsection{SQUID Reaches Low Noise}

\begin{figure}[!p]
    \centering
    \includegraphics[width=\textwidth]{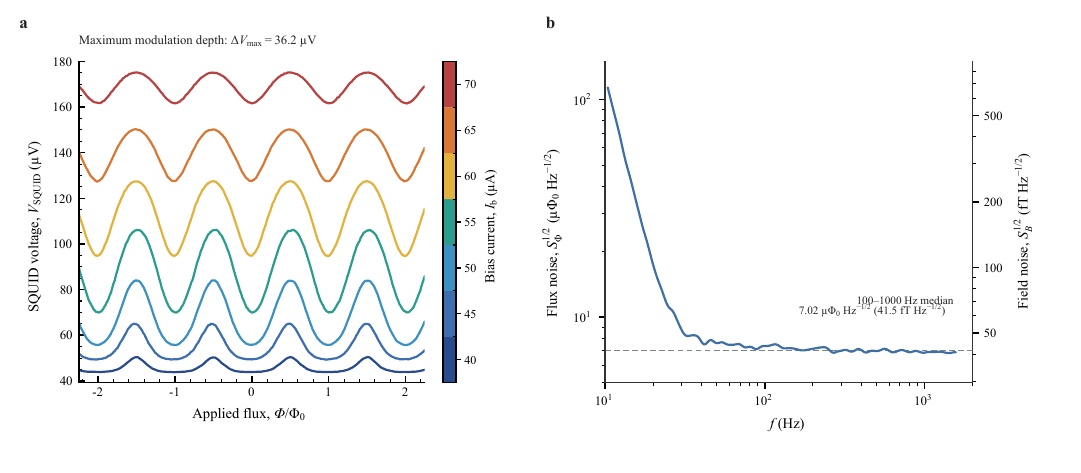}
    \caption{\textbf{SQUID modulation and noise performance.} \textbf{a,} Voltage-flux curves measured at bias currents from 40 to 70 $\mu$A, showing a maximum measured modulation depth of 36.2 $\mu$V. The flux coordinate is normalized to the measured response period. \textbf{b,} Noise spectrum measured at 77 K. The gray dashed line indicates the median noise level from 100 to 1000 Hz: 41.5 fT Hz$^{-1/2}$.}
    \label{fig:4}
\end{figure}

Periodic voltage–flux modulation at 77 K was resolved throughout the tested bias-current range from 40 to 70 $\mu$A (Fig.~\ref{fig:4}a), demonstrating stable flux-to-voltage transduction over a comparatively broad operating window. The modulation amplitude varied systematically with bias current and reached a maximum peak-to-peak value of 36.2 $\mu$V at 55 $\mu$A.

The measured noise spectrum showed a pronounced increase toward low frequencies and approached a substantially flatter level in the higher-frequency region (Fig. 4b). The median flux-noise spectral density between 100 to 1000 Hz was 7.02 $\mu\Phi_0\,\mathrm{Hz}^{-1/2}$, corresponding to a field-noise of 41.5 fT Hz$^{-1/2}$, as obtained using an effective area of 0.35 mm$^{2}$. This performance lies within the state-of-the-art range reported for HTS SQUID magnetometers operated at liquid-nitrogen temperature \cite{Beyer1998SUST,Lam2013,trabaldo2019,Ruffieux2020SUST,foley2021,Xu2023pkubicrystal100fT,}. Importantly, this low-noise behavior is retained when the SQUID is operated in the developed nonmagnetic G10 Dewar, establishing compatibility between the long-duration cryogenic platform and high-sensitivity magnetometry.

\section{Discussion}

The main implication of this work is that the Dewar should be treated as part of the magnetometer architecture rather than as an independent cryogenic container. In a compact HTS-SQUID system, cryogen endurance, magnetic cleanliness, sensor geometry, instrument size, and portability are coupled through the same mechanical design. The design problem is therefore not simply to maximize liquid-nitrogen capacity or minimize heat load, but to identify a practical operating point among competing system-level requirements. This constraint-based view is particularly relevant to compact and portable SQUID instruments, for which increasing vessel size or insulation is not, by itself, a viable route to longer operation.

The thermal model is useful in this context primarily as a design tool rather than as a detailed calorimetric description of the Dewar. Its agreement with the long-duration inventory measurement shows that the overall thermal behavior can be represented with sufficient fidelity to compare candidate geometries and project the usable cryogenic duration beyond the measurement window. The significance of the model is therefore its predictive value within the defined design space. Repeated endurance measurements on additional units will therefore be required in future work to quantify vessel-to-vessel variability and establish their transferability.

A practical SQUID Dewar must ultimately be judged not only by how long it retains cryogen, but also by whether it preserves the performance of the sensor it is intended to support. The low-noise operation demonstrated here indicates that the lightweight nonmagnetic G10 architecture introduces no evident system-level penalty to the HTS-SQUID measurement chain. This links cryogenic endurance and magnetic performance within a single compact platform, and suggests that long-duration operation need not be achieved at the expense of SQUID sensitivity. More broadly, the approach provides a transferable framework for designing cryogenic enclosures in compact HTS magnetometers where thermal, magnetic, geometric, and portability requirements must be satisfied simultaneously.

\section*{Acknowledgments}

This work was supported by the National Key R\&D Program of China (2023YFF0720500) and the Research Startup Funds of Hangzhou International Innovation Institute of Beihang University (2025BKZ030).

\subsection*{Author Contributions}

X. Shi, B. Xiang, and L. Hou jointly contributed to the study design, methodology development, experimental investigation, data analysis, visualization, and manuscript preparation. W. Tang, G. Zhang, S. Liu, R. Wang, Y. Wang, and Z. Cao contributed to the experimental investigation, device-related measurements, and data validation. X. Wang and X. Zhang provided technical guidance and resources. B. Xiang and X. Lin supervised and coordinated the project. All authors discussed the results, reviewed the manuscript, and approved the final version.

\subsection*{Conflicts of Interest}

The authors declare that there is no conflict of interest regarding the publication of this article.

\subsection*{Data Availability}

The data and analysis code supporting this study are available from the corresponding author upon reasonable request.

\subsection*{Ethics Statement}

This study involved no human participants, animals, biospecimens, or identifiable personal data; human- or animal-subject approval was not applicable.

\printbibliography

\end{document}